\documentclass[sigconf,screen]{acmart}

\usepackage{booktabs}
\usepackage{multirow}
\usepackage{makecell}
\usepackage{subfigure}
\usepackage[misc]{ifsym} 
\usepackage{balance}
\usepackage{listings}
\usepackage{color, xcolor}
\usepackage{algorithm2e}
\usepackage{diagbox}
\usepackage{tcolorbox}
\usepackage{stfloats}
\usepackage{enumitem}
\usepackage{cleveref}
\usepackage{ulem}
\usepackage{colortbl} 

\acmYear{2026}\copyrightyear{2026}
\setcopyright{cc}
\setcctype[4.0]{by}
\acmConference[FSE Companion '26]{34th ACM Joint European Software Engineering Conference and Symposium on the Foundations of Software Engineering}{July 5--9, 2026}{Montreal, QC, Canada}
\acmBooktitle{34th ACM Joint European Software Engineering Conference and Symposium on the Foundations of Software Engineering (FSE Companion '26), July 5--9, 2026, Montreal, QC, Canada}
\acmDOI{10.1145/3803437.3805206}
\acmISBN{979-8-4007-2636-1/26/07}

\definecolor{lightcoral}{rgb}{0.94, 0.5, 0.5}
\definecolor{lightgreen}{rgb}{0.56, 0.93, 0.56}
\definecolor{harvestgold}{rgb}{0.85, 0.57, 0.0}
\definecolor{brightlavender}{rgb}{0.75, 0.58, 0.89}
\definecolor{capri}{rgb}{0.0, 0.75, 1.0}
\definecolor{carminepink}{rgb}{0.92, 0.3, 0.26}
\definecolor{celadon}{rgb}{0.67, 0.88, 0.69}
\definecolor{darkpastelgreen}{rgb}{0.01, 0.75, 0.24}
\definecolor{DeepSkyBlue4}{RGB}{0,104,139}

\definecolor{acccolor}{RGB}{238, 245, 252} 
\definecolor{latcolor}{RGB}{242, 242, 242} 

\crefformat{section}{\S#2#1#3} 
\crefformat{subsection}{\S#2#1#3}
\crefformat{subsubsection}{\S#2#1#3}

\begin{document}
\begin{sloppypar}
	
	\author{Lingzhe Zhang$^{\dag}$}
	\affiliation{%
		\institution{Peking University; Key Laboratory of Data Intelligence and Security}
		\city{Beijing}
		\country{China}}
	\email{zhang.lingzhe@stu.pku.edu.cn}
	
	\author{Yunpeng Zhai$^{\dag}$}
	\thanks{$^{\dag}$Equal contribution}
	\affiliation{%
		\institution{Alibaba Group}
		\country{China}}
	\email{zhaiyunpeng.zyp@alibaba-inc.com}
	
	\author{Tong Jia$^{\ast}$}
	\thanks{$^{\ast}$Corresponding author}
	\affiliation{%
		\institution{Peking University; Key Laboratory of Data Space Technology and System}
		\city{Beijing}
		\country{China}}
	\email{jia.tong@pku.edu.cn}
	
	\author{Minghua He}
	\affiliation{%
		\institution{Peking University; Key Laboratory of Data Intelligence and Security}
		\city{Beijing}
		\country{China}}
	\email{hemh2120@stu.pku.edu.cn}
	
	\author{Chiming Duan}
	\affiliation{%
		\institution{Peking University; Key Laboratory of Data Intelligence and Security}
		\city{Beijing}
		\country{China}}
	\email{duanchiming@stu.pku.edu.cn}
	
	\author{Zhaoyang Liu}
	\affiliation{%
		\institution{Alibaba Group}
		\country{China}}
	\email{jingmu.lzy@alibaba-inc.com}
	
	\author{Bolin Ding}
	\affiliation{%
		\institution{Alibaba Group}
		\country{China}}
	\email{bolin.ding@alibaba-inc.com}
	
	\author{Ying Li$^{\ast}$}
	\affiliation{%
		\institution{Peking University; Key Laboratory of Data Intelligence and Security}
		\city{Beijing}
		\country{China}}
	\email{li.ying@pku.edu.cn}
	
\renewcommand{\shortauthors}{Lingzhe Zhang et al.}

\title[E2E-REME]{E2E-REME: Towards End-to-End Microservices Auto-Remediation via Experience-Simulation Reinforcement Fine-Tuning}

\begin{abstract}
	Contemporary microservice systems continue to grow in scale and complexity, leading to increasingly frequent and costly failures. While recent LLM-based auto-remediation approaches have emerged, they primarily translate textual instructions into executable Ansible playbooks and rely on expert-crafted prompts, lacking runtime knowledge guidance and depending on large-scale general-purpose LLMs, which limits their accuracy and efficiency. We introduce \textit{End-to-End Microservice Remediation} (E2E-MR), a new task that requires directly generating executable playbooks from diagnosis reports to autonomously restore faulty systems. To enable rigorous evaluation, we build \textit{MicroRemed}, a benchmark that automates microservice deployment, failure injection, playbook execution, and post-repair verification. We further propose \textit{E2E-REME}, an end-to-end auto-remediation model trained via experience-simulation reinforcement fine-tuning. Experiments on public and industrial microservice platforms, compared with nine representative LLMs, show that E2E-REME achieves superior accuracy and efficiency.
\end{abstract}

\ccsdesc[500]{Software and its engineering~Maintaining software}

\keywords{Auto-Remediation, Reinforcement Fine-Tuning, Microservices}

\maketitle

\section{Introduction}

Modern microservice systems have become increasingly complex due to dynamic interactions and rapidly evolving runtime environments~\cite{zhang2024failure}. This rising complexity inevitably leads to more frequent and harder-to-predict system failures. Such failures can be extremely costly: according to industry analysis, large enterprises experience an average direct loss of exceed \$1,000,000 for every hour of IT downtime~\cite{ITICCostofDowntime2024}.

Given the substantial operational and financial impact, modern cloud-native systems urgently require the capability not only to detect failures but also to automatically remediate them in a timely and reliable manner~\cite{remil2024aiops, duan2025logaction, he2025walk, liu2025ora, zhang2024reducing, zhang2024time}. As a result, auto-remediation—which autonomously identifies appropriate corrective actions and executes them with minimal human intervention—has become a critical component for ensuring resilient and cost-efficient operations at scale~\cite{zhang2025log, he2025united, zhang2026hypothesize, zhang2026agentic, huang2025uda, liu2025aaad, zhang2026runtimeslicer, zhang2026efficient, kang2022separation}.

With the rapid advancement of large language models (LLMs), researchers have increasingly explored leveraging their strong reasoning and code-generation capabilities~\cite{zhang2025surveyagents, zhang2025survey2, pan2025omni, joel2024survey, singh2025agentic, zhang2025agentfm, zhang2025thinkfl, zhang2025adaptive, pan2025d, hong2025cslparser} for microservice remediation~\cite{zhang2025survey}. A practical and industry-adoptable approach to address microservice auto-remediation is to use LLMs to generate ansible playbooks that can be automatically executed to repair faulty services~\cite{sahoo2024ansible}. Compared with traditional shell scripts, Ansible playbooks serve as structured, declarative specifications for operational procedures and offer a higher-level abstraction with clearer structure, stronger readability, and improved reusability and maintainability~\cite{hochstein2017ansible}. These advantages have made playbooks a widely adopted mechanism for implementing automated operational procedures in large-scale microservice systems, and thus a natural target for LLM-driven auto-remediation~\cite{sarda2024leveraging}.

Following this trend, existing work on LLM-driven microservice remediation can be broadly categorized into two groups: methods and benchmarks. Methods focus on generating executable ansible playbooks from human-written instructions. For example, Wisdom-Ansible~\cite{pujar2023automated} fine-tunes CodeGen to produce remediation scripts, MAPE-Ansible~\cite{sarda2024leveraging} leverages GPT-4 and LLaMa-2 70B in a MAPE-K loop architecture, and WCA-Ansible~\cite{sahoo2024ansible} is pre-trained from scratch on natural language, source code, and Ansible data. Benchmarks support these studies by providing curated collections of prompts and playbook templates for automation tasks. KubePlaybook~\cite{namrud2024kubeplaybook} contains 130 natural language prompts for generating automation-focused remediation scripts, while Andromeda~\cite{opdebeeck2021andromeda} provides structural representations of over 125,000 Ansible roles, along with more than 800,000 concrete changes between role versions extracted from the underlying Git repositories. However, existing methods and benchmarks still face key limitations when applied in real-world scenarios:

\begin{itemize}[leftmargin=*]
	\item \textbf{Task-level}: These approaches typically rely on human-crafted prompts authored by experienced SREs, where the LLM only translates textual instructions into executable code. Such designs depend heavily on manual intervention, lack iterative feedback from the runtime environment, and fail to achieve end-to-end automation from failure diagnosis to system recovery.
	\item \textbf{Method-level}: The generation of Ansible playbooks critically depends on the current runtime state of the microservice system. Without accurate and up-to-date system state guidance, the generated remediation scripts may be suboptimal or even incorrect. Moreover, representative methods such as MAPE-Ansible~\cite{sarda2024leveraging} rely on very large, closed-source models (e.g., GPT-4 and LLaMa-2 70B), which require substantial computational resources and time for inference, limiting both scalability and efficiency.
\end{itemize}

\begin{figure}[htbp]
	\centering
	\includegraphics[width=1\linewidth]{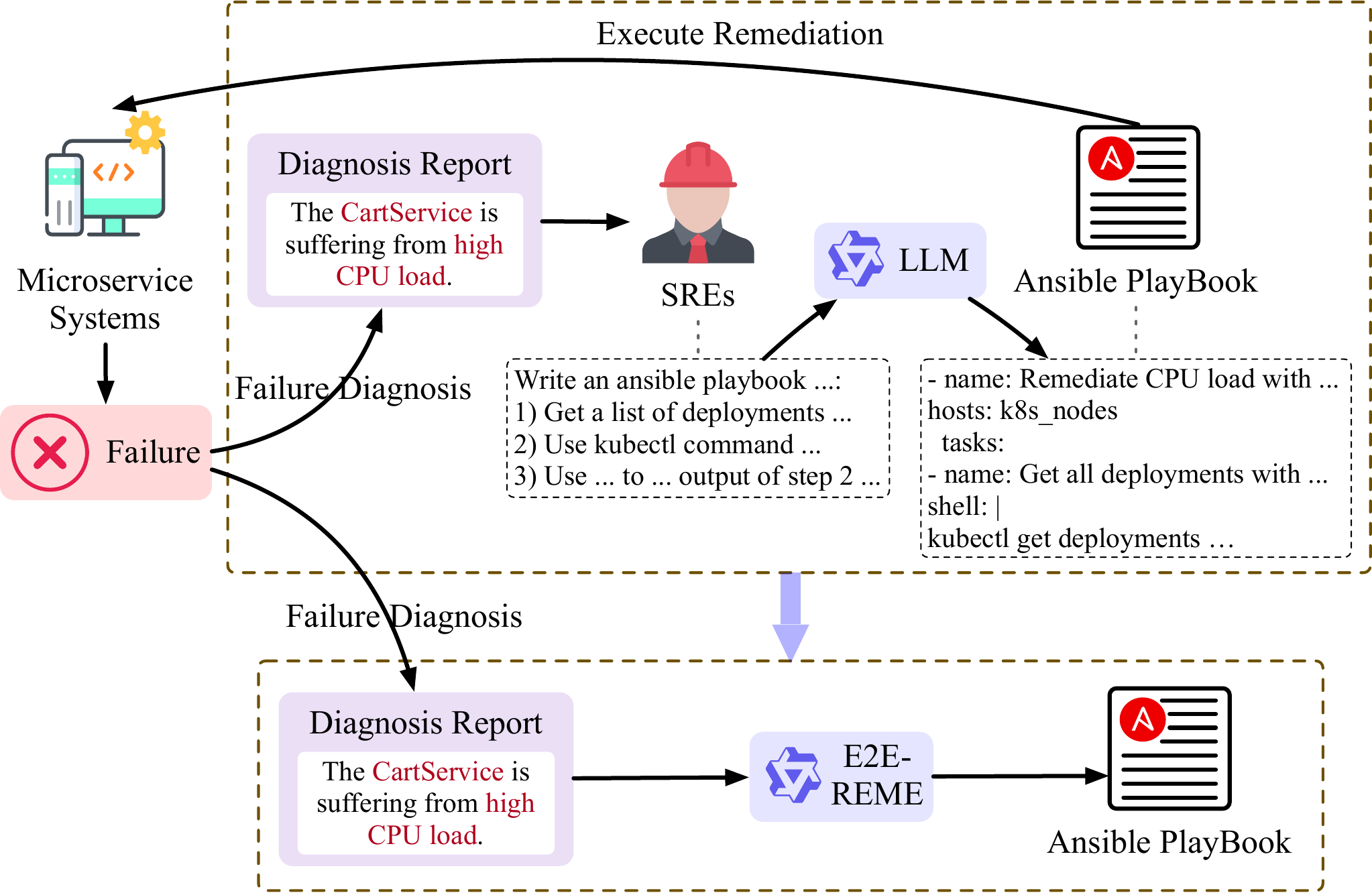}
	\caption{Previous microservice remediation workflow compared with the end-to-end microservice remediation pipeline proposed in this paper.}
	\label{fig: intro-example}
	\vspace{-1em}
\end{figure}

\uline{To address the task-level challenge, we propose a new task, End-to-End Microservice Remediation (E2E-MR)}. E2E-MR aims to directly generate executable ansible playbooks from a given diagnosis report and autonomously recover the faulty system. As illustrated in Figure~\ref{fig: intro-example}, unlike previous approaches that rely on human-crafted prompts authored by experienced SREs based on diagnostic reports, E2E-MR establishes a closed-loop remediation pipeline, in which LLMs translate diagnostic insights into concrete repair actions that can be automatically executed within the microservice environment.

For evaluation and structured comparison of this challenging new task, we introduce \textit{MicroRemed}\footnote{The benchmark is available at \url{https://github.com/LLM4AIOps/MicroRemed}.}, a benchmark designed to assess LLMs’ capabilities in end-to-end microservice remediation. MicroRemed automatically deploys a real microservice system and continuously injects diverse failures. For each injected failure, it generates a corresponding diagnosis report based on the target component and failure type, which is then provided to the LLM under evaluation. The LLM produces an Ansible playbook, which is executed automatically, and the system subsequently verifies whether the injected failure has been successfully repaired. MicroRemed supports unlimited rounds of random failure injection and verification, allowing for extensive stress testing and iterative evaluation. Moreover, to facilitate fair and structured comparison, we categorize remediation targets into three difficulty levels—easy, medium, and hard—based on the complexity and interdependency of the underlying failure scenarios.

\uline{To address the method-level challenges, we propose E2E-REME, an end-to-end auto-remediation model for microservices via Experience-Simulation Reinforcement Training}. \textbf{E2E-REME}\footnote{The model is available at \url{https://modelscope.cn/models/ZhangLingzhe/E2E-REME}.} is designed to mirror how human SREs handle failures: continuously acquiring fresh runtime signals, reasoning over candidate repair strategies, and refining their decisions before executing a final remediation plan. To operationalize this process, E2E-REME employs a lightweight multi-agent workflow, which we refer to as ThinkRemed. It serves as an architectural scaffold that structures remediation into probing, executing, and verifying-refinement steps.

Training E2E-REME for the end-to-end microservice remediation task is accomplished via an \textit{Experience-Simulation Reinforcement Training} pipeline tailored to microservice operations. The pipeline consists of three stages: (1) Expert-Guided Supervised Fine-Tuning (SFT), which provides foundational remediation behaviors; (2) Simulation-Based RFT, which exposes the model to synthetic failure scenarios and teaches it to act within simulated environments; and (3) Reality-Anchored RFT, which further optimizes the model using feedback collected from real remediations.

We evaluate E2E-REME using MicroRemed, integrated with two widely adopted microservice systems—Train-Ticket~\cite{zhou2018fault} and Online-Boutique~\cite{google2025onlineboutique}—as well as a self-developed lightweight system, Simple-Micro. Experimental results show that E2E-REME surpasses nine representative LLMs, achieving up to an average 49.32\% higher accuracy while maintaining competitive inference efficiency. We further validate its performance under realistic industrial workloads and microservice environments, which similarly confirm the effectiveness and robustness of E2E-REME.

In summary, the contributions of our work are as follows:
\begin{itemize}[itemsep=1pt, topsep=2pt, parsep=0pt, partopsep=0pt, leftmargin=*]
	\item We introduce the task of end-to-end microservice remediation (E2E-MR), which requires LLMs to directly generate executable Ansible playbooks from diagnosis reports and autonomously repair faulty systems. To support systematic evaluation, we build \textbf{MicroRemed}, a challenging benchmark that automates microservice deployment, failure injection, playbook execution, and post-repair verification.
	\item To address the challenges of E2E-MR, we propose E2E-REME, an end-to-end auto-remediation model for microservices. E2E-REME operates within a lightweight multi-agent workflow, ThinkRemed, which structures the remediation process into probing, execution, and verification–refinement steps. The model is trained via a tailored Experience-Simulation Reinforcement Training pipeline.
	\item We conduct extensive experiments on three microservice systems and compare E2E-REME against nine representative LLMs. Results demonstrate its superior accuracy and efficiency. Additional validation under realistic industrial-style workloads further confirms the effectiveness and robustness of E2E-REME.
\end{itemize}

\section{Background}

\subsection{Microservice Auto-Remediation}

Modern microservice architectures decompose applications into large numbers of loosely coupled services that interact through lightweight APIs. While this design improves scalability and development agility, it also increases operational complexity: faults may arise from configuration errors, resource contention, cascading failures, or inconsistent service states. As a result, automated remediation has become essential for maintaining system reliability.

Auto-remediation refers to the process of automatically generating a repair plan, and executing the necessary actions to restore system health. In practice, these repair actions often involve operations such as restarting services, modifying configurations, cleaning corrupted state, or redeploying components. Ansible is widely used for such tasks because it provides a declarative automation framework that can execute system-level and application-level operations across distributed environments. Its playbook-based design enables LLMs or automation agents to produce actionable repair procedures that can be directly executed without human intervention.

An Ansible playbook is a YAML-based script defining hosts, tasks, and their execution conditions. Figure~\ref{fig: ansible-playbook} is a simplified example illustrating how a remediation workflow can handle high CPU load by automatically scaling a service.

\lstset{
	basicstyle=\ttfamily\footnotesize,
	backgroundcolor=\color{gray!10},
	frame=single,
	rulecolor=\color{black},
	frameround=tttt,
	showstringspaces=false,
	breaklines=true,
	postbreak=\mbox{\textcolor{red}{$\hookrightarrow$}\space},
	numbers=left,
	numberstyle=\tiny\color{gray},
	xleftmargin=1em,
	xrightmargin=1em,
	keywordstyle=\color{blue},
	stringstyle=\color{teal},
}

\begin{figure}[htbp]
	\centering
	\begin{lstlisting}
---
- name: Mitigate high CPU load
hosts: microservice_nodes
become: yes
tasks:
- name: Check CPU usage
shell: "top -bn1 | awk -F'[, ]+' '/Cpu/{print $3+$5}'"
register: cpu

- name: Scale service if CPU > 80%
shell: kubectl scale deploy my-service --replicas=4
when: cpu.stdout | float > 80

- name: Notify monitoring
shell: "curl http://monitor/api/notify -d 'scaled'"
	\end{lstlisting}
	\vspace{-1em}
	\caption{An Ansible Playbook for CPU scaling}
	\label{fig: ansible-playbook}
	\vspace{-1em}
\end{figure}

This example shows how Ansible playbook operationalizes auto-remediation by providing a structured, declarative interface that bridges diagnosis and repair—allowing LLM-based systems to generate actionable and directly executable recovery procedures.

\subsection{Reinforcement Fine-Tuning}

Reinforcement Fine-Tuning (RFT) adapts language models to decision-making tasks by optimizing model behavior using reward or preference feedback, rather than relying solely on supervised instruction–response pairs~\cite{christiano2017deep, ziegler2019fine}. Unlike supervised fine-tuning (SFT), which teaches models to imitate expert demonstrations, RFT enables models to explore action spaces, evaluate long-term consequences, and self-correct through iterative interaction with an environment. RFT methods can be broadly categorized into two families:

\textbf{(1) Reward-based policy optimization.}
These approaches assign scalar rewards to model-generated actions or trajectories and optimize the policy to maximize expected reward. Group-based variants, such as Group Relative Policy Optimization (GRPO)~\cite{shao2024deepseekmath}, enhance stability by comparing multiple model completions under the same context, enabling the model to learn nuanced distinctions among candidate actions. Such methods are particularly effective for structured reasoning and tool-use scenarios where reward signals derive from execution correctness, efficiency, or safety.

\textbf{(2) Preference-based optimization.}
In many real-world decision-making tasks, explicit numeric rewards are difficult to define, while human operators can reliably express preferences between paired model outputs. Direct Preference Optimization (DPO)~\cite{rafailov2023direct} provides a scalable solution by learning directly from such pairwise preferences, aligning model behavior with human judgments without requiring reinforcement learning rollouts or handcrafted reward models.

Both families offer complementary strengths: reward-based methods facilitate exploration and environment-driven learning, whereas preference-based methods enable fine-grained alignment with human operational expertise.

Despite rapid progress, the application of RFT to system operations—particularly microservice auto-remediation—remains limited. Compared with static preference or tool-use settings, auto-remediation introduces uniquely challenging characteristics: dynamic runtime states, multi-step causal dependencies, safety-critical actions, and scarce human supervision. These factors motivate a combined RFT strategy in E2E-REME, leveraging reward-driven exploration in simulated environments together with preference-driven alignment using real-world operator corrections.

\begin{figure*}[tbp]
	\centering
	\includegraphics[width=1\linewidth]{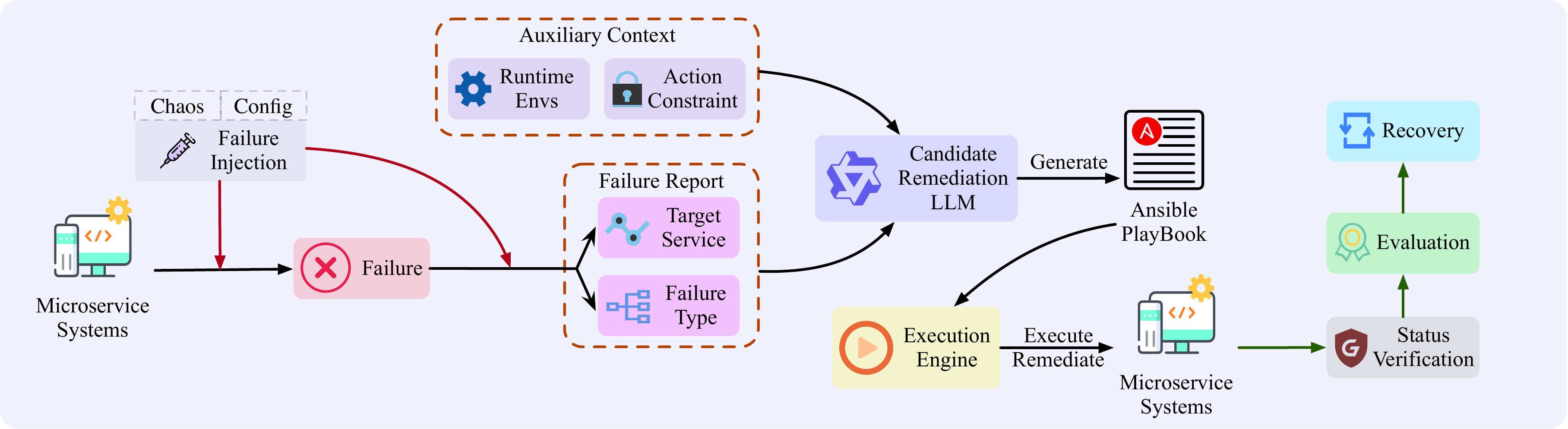}
	\caption{MicroRemed Benchmark Pipeline: the benchmark launches a real microservice; Failure Injection injects faults and produces a Failure Report; the Failure Report together with Auxiliary Context is provided to the Candidate Remediation LLM which generates an Ansible Playbook; the Execution Engine executes the playbook; Status Verification checks remediation success; Evaluation and Recovery restores the system for the next run.}
	\label{fig: benchmark}
\end{figure*}

\section{Benchmark Construction}

We present the construction of the MicroRemed benchmark in this section. We begin with an overview of the task definition and the underlying design principles (\cref{sec: design-principles}), followed by the architecture of the MicroRemed benchmark (\cref{sec: architecture}) and the evaluation protocol (\cref{sec: evaluation-protocol}). Finally, we describe the overall composition of MicroRemed (\cref{sec: benchmark-composition}).

\subsection{Design Principles}
\label{sec: design-principles}

Existing microservice remediation approaches typically depend on human-crafted prompts designed by experienced SREs, where LLMs merely translate natural language instructions into executable scripts such as Ansible playbooks. This paradigm lacks autonomy and generalization, as it relies heavily on explicit human reasoning rather than the model’s understanding of the system state.

To address this limitation, we introduce the task of \textbf{End-to-End Microservice Remediation (E2E-MR)}, which aims to evaluate an LLM’s ability to autonomously generate executable remediation plans given only structured diagnostic information. Unlike conventional prompt-based generation, E2E-MR emphasizes a direct remediation process that transforms diagnostic reports into actionable repair operations.

\begin{equation}
	\left\{
	\begin{aligned}
		f_\theta &: (\mathcal{S}_{target}, \mathcal{T}_{fail}, \mathcal{C}_{aux}) \to p^*,\\
		p^* &= \arg\max_{p \in \mathcal{P}} 
		\mathcal{U}\big(\mathcal{E}(p, \mathcal{S}_{fail}) = \mathcal{S}_{normal}\big)
	\end{aligned}
	\right.
	\label{eq: e2e-mr}
\end{equation}

Formally, the E2E-MR task can be formulated as Equation~\ref{eq: e2e-mr}, where $f_\theta$ is the candidate remediation LLM parameterized by $\theta$, $\mathcal{S}_{target}$ denotes the failed microservice, $\mathcal{T}_{fail}$ the failure type, and $\mathcal{C}_{aux}$ auxiliary contextual information. $\mathcal{P}$ is the space of executable playbooks, $\mathcal{E}$ represents the execution environment, and $\mathcal{U}(\cdot)$ measures the utility of successful recovery. The goal is to generate an optimal playbook $p^*$ that maximizes the likelihood of recovering the system state $\mathcal{S}_{fail}$ to $\mathcal{S}_{normal}$.

Therefore, to design a benchmark for the E2E-MR task, we adhere to the following design principles:

\begin{itemize}[leftmargin=*]
	\item \textbf{Dynamic Execution Benchmark.}
	Unlike most LLM benchmarks that collect static data to form fixed datasets, the proposed benchmark is designed as a live and interactive execution environment. It actively launches real microservice systems, injects controlled failures, and interacts dynamically with running services. This design enables the benchmark to capture real-time behaviors, system dynamics, and contextual dependencies that static datasets cannot represent.
	\item \textbf{Execution-based Evaluation.}  
	Evaluation is not determined by linguistic or structural similarity of generated outputs, but by execution outcomes. Each generated playbook is executed within the microservice environment, and the benchmark verifies success by assessing whether the system has been fully recovered to its normal operational state.
	\item \textbf{Comprehensive Scalability.}  
	Built on these foundations, the benchmark is designed to be method-scalable, LLM-scalable, failure-scalable, and system-scalable. It supports diverse LLM-based remediation methods, allows plug-and-play replacement of remediation models, accommodates various failure scenarios, and can be easily extended to new microservice systems with minimal configuration effort.
\end{itemize}

\begin{figure*}[h]
	\centering
	\includegraphics[width=1\linewidth]{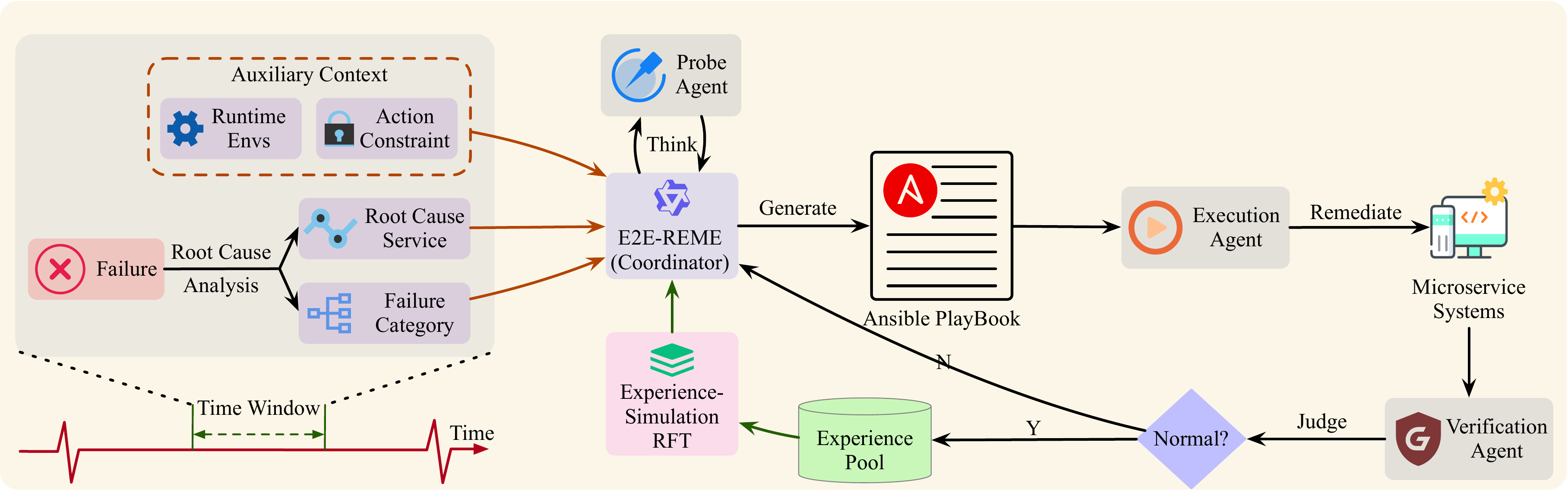}
	\caption{Runtime pipeline of E2E-REME. The model acts as a coordinator within a multi-agent workflow, \textit{ThinkRemed}, which organizes the remediation process through the coordination of probing, execution, and verification agents.}
	\label{fig: pipeline}
\end{figure*}

\subsection{Architecture}
\label{sec: architecture}

Based on the above design principles, we develop MicroRemed. The overall architecture of MicroRemed is illustrated in Figure~\ref{fig: benchmark}. MicroRemed actively launches real microservice systems and performs Failure Injection to introduce controlled faults. According to the injected target service and failure type, it generates a Failure Report, which—together with a set of Auxiliary Contexts—is provided to the Candidate Remediation LLM to produce an executable Ansible Playbook. The playbook is then executed by an Execution Engine to carry out automated remediation. After execution, a Status Cerification module checks whether the issue has been successfully resolved.
Finally, the Evaluation and Recovery stage assesses the remediation outcome and restores the microservice system to its original state, thereby enabling reproducible and iterative experimentation.

The Failure Injection module introduces faults into the system through two complementary approaches: chaos injection and configuration injection. For resource-related or runtime failures (e.g., CPU stress, memory pressure, or network latency), MicroRemed adopts chaos injection, which dynamically perturbs the runtime environment using Chaos Mesh~\cite{mesh2025powerful} to emulate realistic fault conditions. For configuration-related failures (e.g., incorrect environment variables or service dependency misconfigurations), the system applies configuration injection, which directly modifies specific configuration files or environment settings to trigger controlled failures.

The Status Verification module resembles traditional anomaly detection in purpose but differs fundamentally in mechanism. While anomaly detection infers abnormality from large volumes of complex runtime data, status verification performs targeted validation of whether a specific injected failure has been fully remediated. For example, if a CPU-stress failure was injected into service A, status verification will exclusively inspect the CPU metrics of service A to confirm recovery. This targeted design ensures 100\% verification accuracy, a level of precision unattainable by general anomaly detection approaches.

\subsection{Evaluation Protocol}
\label{sec: evaluation-protocol}

MicroRemed supports comprehensive evaluation from multiple perspectives, including performance, efficiency, and resource utilization. Specifically, we adopt the following metrics to quantify the effectiveness of candidate remediation LLMs:

\textbf{Remediation Accuracy (\textit{RA})} — measures the proportion of failures that are successfully repaired, reflecting the overall performance of the model.

\textbf{Average Remediation Latency (\textit{ARL})} — evaluates the temporal efficiency of each successful remediation cycle, encompassing both reasoning and execution delays.

\textbf{Average Token Consumption (\textit{ATC})} — quantifies the language-model cost efficiency, representing the average number of tokens consumed to achieve a successful remediation.

\subsection{Benchmark Composition}
\label{sec: benchmark-composition}

Although MicroRemed is designed with comprehensive scalability and supports extensible failure types and microservice systems, in our benchmark we include seven representative types of failures and three real-world microservice systems.

\textbf{Failure Types.} As shown in Table~\ref{tab: failure-types}, MicroRemed includes seven representative failures across three categories: resource-level (CPU, memory, I/O saturation), network-level (network loss, network delay), and application-level (pod failure, configuration error).

\begin{table}[htb]
	\setlength{\tabcolsep}{10pt}
	\centering
	\begin{tabular}{ccc}
		\toprule
		\textbf{No.} & \textbf{Category} & \textbf{Failure Types} \\
		\midrule
		1 & \multirow{3}*{Resource-Level} & CPU Saturation \\
		2 & ~ & Memory Saturation \\
		3 & ~ & IO Saturation \\
		\midrule
		4 & \multirow{2}*{Network-Level} & Network Loss \\
		5 & ~ & Network Delay \\
		\midrule
		6 & \multirow{2}*{Application-Level} & Pod Failure \\
		7 & ~ & Configuration Error \\
		\bottomrule
	\end{tabular}
	\vspace{0.5em}
	\caption{Benchmark statistics on failure types}
	\label{tab: failure-types}
	\vspace{-2em}
\end{table}

\textbf{Microservice Systems.} MicroRemed integrates three microservice systems. Among them, two widely used benchmarks—Train-Ticket~\cite{zhou2018fault} and Online-Boutique~\cite{google2025onlineboutique}—are well recognized for emulating realistic production environments. In addition, we include a self-developed lightweight system, Simple-Micro, designed to enable controlled experiments and facilitate fine-grained analysis.

\textbf{Difficulty Levels.} Although MicroRemed supports arbitrary combinations of injected failures, we define three standardized difficulty levels—easy (23 cases), medium (49 cases), and hard (80 cases)—to enable fair and structured comparison across remediation methods. Each level corresponds to a curated set of failure combinations that vary in fault diversity, dependency complexity, and recovery difficulty.

\section{E2E-REME}

To address the end-to-end microservice auto-remediation task, we propose E2E-REME, a model designed to emulate how human SREs diagnose and repair failures. As shown in Figure~\ref{fig: pipeline}, E2E-REME operates by continuously gathering fresh runtime signals, reasoning over potential repair strategies, and iteratively refining decisions before executing a final remediation plan. To operationalize this workflow, E2E-REME adopts a lightweight multi-agent framework, \textbf{ThinkRemed}, which structures the remediation process into three agents: probing (collecting runtime evidence), execution (applying candidate repairs), and verification–refinement (validating and adjusting actions based on system feedback). Built on top of this framework, we further train E2E-REME using a tailored \textbf{Experience-Simulation Reinforcement Training (RFT)} pipeline, enabling the model to learn robust, action-oriented remediation behaviors aligned with real microservice operational dynamics.

\subsection{ThinkRemed: A Multi-Agent Microservice Auto-Remediation Framework}

Figure~\ref{fig: pipeline} illustrates the operational workflow of \textbf{ThinkRemed}. When a microservice system experiences a failure, a state-of-the-art root cause analysis identifies the faulty service and the corresponding failure category. This information, together with auxiliary context (e.g., runtime environment and action constraints), is provided as input to E2E-REME, which acts as the \textit{Coordinator} within ThinkRemed. 

The \textit{Coordinator} first receives the auxiliary context $\mathcal{C}_0$ and failure report $\mathcal{R}_0$, and adaptively determines whether to invoke the \textit{Probe Agent} to gather additional runtime information from the system. The probe agent executes a series of system state queries and returns the corresponding results. Once sufficient information is collected, the \textit{Coordinator} synthesizes a candidate Ansible playbook $p_t$.

The generated playbook $p_t$ is then sent to the \textit{Execution Agent}, which attempts to remediate the faulty microservice system and records the execution outcomes for subsequent reflection. After execution, the \textit{Verification Agent} evaluates the remediation result, producing a binary outcome $v_t \in \{0,1\}$ indicating success or failure. It is important to note that this \textit{Verification Agent} differs from the \textit{Status Verification} used in the benchmark. In the benchmark’s simulated failure environment, \textit{Status Verification} can directly access low-level system information to determine the repair outcome. In contrast, the \textit{Verification Agent} operates in a live system setting and relies on state-of-the-art anomaly detection methods to assess whether the remediation was successful. If the remediation fails, the system enters a reflection phase, and control returns to the \textit{Coordinator} for iterative refinement based on the feedback. To ensure timely remediation and accommodate LLM context limitations, the iteration loop is bounded by a maximum trial budget $T_{\max}$.

\begin{equation}
	\left\{
	\begin{aligned}
		p_t &= f_\theta(\mathcal{R}_t, \mathcal{C}_t, \mathcal{I}_t), \\
		s_{t+1} &= \mathcal{E}(p_t, s_t), \\
		v_t &= \mathcal{V}(s_{t+1}), \\
		(\mathcal{R}_{t+1}, \mathcal{C}_{t+1}) &= \mathcal{U}(\mathcal{R}_t, \mathcal{C}_t, s_{t+1}) \\
		&\qquad \text{if } v_t = 0 \text{ and } t < T_{\max}
	\end{aligned}
	\right.
	\label{eq: think-remed}
\end{equation}

In summary, the iterative process of ThinkRemed can be formalized as Equation~\ref{eq: think-remed}, where $f_\theta$ denotes the Coordinator’s reasoning policy, $\mathcal{E}$ the execution operator, and $\mathcal{V}$ the verification predicate. This multi-agent, iterative workflow enables E2E-REME to continuously reason, act, and refine its remediation strategies in response to dynamic microservice system states.

\subsection{Experience-Simulation RFT}

Experience-Simulation RFT consists of three stages: (1) Expert-Guided SFT, (2) Simulation-Based RFT, and (3) Reality-Anchored RFT. The first two stages—Expert-Guided SFT and Simulation-Based RFT—are used to train E2E-REME in an offline setting, while Reality-Anchored RFT continuously fine-tunes the model after it is deployed online.

\begin{figure}[tbp]
	\centering
	\includegraphics[width=1\linewidth]{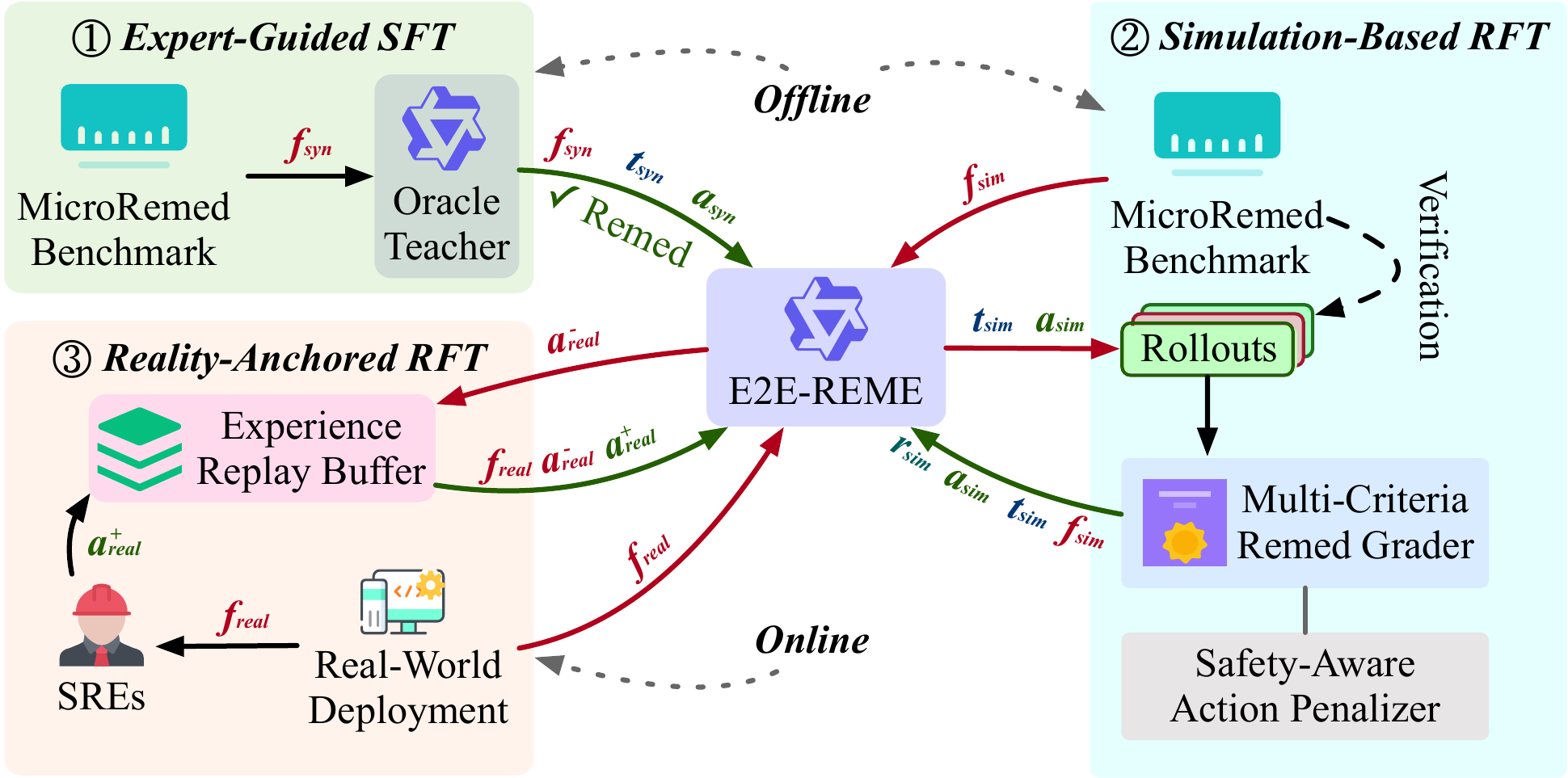}
	\caption{Overall framework of \textbf{Experience-Simulation RFT}}
	\label{fig: ES-RFT}
\end{figure}

\subsubsection{Expert-Guided SFT}

Given that lightweight models exhibit limited reasoning ability, weak tool-use proficiency, and difficulty in directly generating executable Ansible playbooks, the goal of this stage is to teach the model the fundamentals of (i) structured reasoning, (ii) tool-calling behaviors, and (iii) valid Ansible playbook construction.

To achieve this, we employ an Oracle Teacher Model and run it on the MicroRemed Benchmark. For each synthetic failure instance $f_{syn}$, the teacher model is allowed to interact with the environment. If the remediation attempt succeeds, we extract the reasoning trace $t_{syn}$, and the final executable Ansible playbook $a_{syn}$. These elements, paired with the original failure description, form the supervision tuples $\{f_{syn}, t_{syn}, a_{syn}\}$.

\begin{equation}
	\mathcal{L}_{\text{SFT}}
	= \mathbb{E}_{(f_{\text{syn}}, t_{\text{syn}}, a_{\text{syn}})} 
	\left[ 
	- \log \pi_{\theta}(t_{\text{syn}}, a_{\text{syn}} \mid f_{\text{syn}}) 
	\right]
	\label{eq:sft-loss}
\end{equation}

The Expert-Guided SFT phase thus optimizes the model to reproduce both the teacher’s reasoning process and its generated repair actions. Formally, the SFT objective can be expressed as Equation~\ref{eq:sft-loss}, where denotes the policy of E2E-REME.

\subsubsection{Simulation-Based RFT}

While SFT enables the model to imitate expert behaviors and acquire essential formatting and tool-use patterns, it cannot teach the model to reason, explore, or self-correct beyond the expert demonstrations. To equip E2E-REME with these capabilities, we design a Simulation-Based RFT stage.

In this stage, the SFT-initialized E2E-REME is deployed on the MicroRemed benchmark to generate full rollouts. For each simulated episode, the model produces a reasoning trace $t_{\mathrm{sim}}$, an Ansible playbook $a_{\mathrm{sim}}$, and receives a remediation outcome indicating whether the system has been successfully repaired. A reward is then assigned to the rollout using a combination of (1) a \textit{Multi-Criteria Remediation Grader} and (2) a \textit{Safety-Aware Action Penalizer}.

The Multi-Criteria Remediation Grader provides the primary reward signal. It assigns a high reward for successful remediation and further incorporates several auxiliary criteria, including: structural correctness of the generated playbook, successful execution of individual tasks, absence of execution errors, and token-efficiency of the reasoning trace (encouraging concise reasoning when remediation succeeds). 

Complementing this, the Safety-Aware Action Penalizer imposes substantial penalties for unsafe actions. Even though action constraints are provided as auxiliary context, lightweight models occasionally generate playbooks that may lead to harmful system-wide side effects. To prevent such behaviors, any playbook containing unsafe operations receives a large negative penalty, regardless of whether the remediation happens to succeed.

\begin{equation}
	R \;=\; 
	\alpha \cdot \mathbb{I}[\text{success}] 
	\;+\; \beta \cdot r_{\mathrm{struct}}
	\;+\; \gamma \cdot r_{\mathrm{exec}}
	\;+\; \delta \cdot r_{\mathrm{eff}}
	\;-\; \lambda \cdot \mathbb{I}[\text{unsafe}]
	\label{eq:sim_reward}
\end{equation}

In summary, the overall reward for a simulated rollout can be defined as Equation~\ref{eq:sim_reward}, where $\mathbb{I}[\cdot]$ is the indicator function, $r_{\mathrm{struct}}$ measures playbook structural validity, $r_{\mathrm{exec}}$ evaluates execution correctness, $r_{\mathrm{eff}}$ encourages concise reasoning, and $\mathbb{I}[\text{unsafe}]$ flags unsafe or system-risky actions. The coefficients $\alpha,\beta,\gamma,\delta,\lambda$ control the relative influence of each signal.

To optimize E2E-REME under this reward model, we adopt \textit{Group Relative Policy Optimization} (GRPO), which stabilizes credit assignment across long reasoning–action traces. Given a batch of rollouts $\{(f_i, a_i, R_i)\}$, GRPO updates the policy by maximizing advantage-weighted likelihood ratios within each rollout group.

Formally, the optimization objective of this stage is given in Equation~\ref{eq:sim_grpo}. Here, $\pi_\theta$ denotes the parametrized policy of E2E-REME after SFT initialization,   $\mathcal{G}(i)$ denotes the group of rollouts originating from the same simulated failure instance, and the term in parentheses introduces a group-wise baseline to reduce gradient variance and stabilize training.

\begin{equation}
	\mathcal{L}_{\mathrm{GRPO}}
	=
	- \sum_{i}
	\log \pi_\theta(a_i \mid f_i)
	\cdot
	\left(
	R_i 
	- \frac{1}{|\mathcal{G}(i)|}
	\sum_{j \in \mathcal{G}(i)} R_j
	\right)
	\label{eq:sim_grpo}
\end{equation}

Through Simulation-Based RFT, E2E-REME learns to explore, self-correct, and refine its reasoning strategies beyond expert demonstrations, enabling more robust and autonomous remediation behaviors.

\begin{table*}[!t]
	\centering
	\setlength{\tabcolsep}{2.05pt}
	\renewcommand{\arraystretch}{1.2} 
	
	\begin{tabular}{c | >{\columncolor{acccolor}}c >{\columncolor{acccolor}}c >{\columncolor{acccolor}}c >{\columncolor{latcolor}}c >{\columncolor{latcolor}}c >{\columncolor{latcolor}}c | >{\columncolor{acccolor}}c >{\columncolor{acccolor}}c >{\columncolor{acccolor}}c >{\columncolor{latcolor}}c >{\columncolor{latcolor}}c >{\columncolor{latcolor}}c | >{\columncolor{acccolor}}c >{\columncolor{acccolor}}c >{\columncolor{acccolor}}c >{\columncolor{latcolor}}c >{\columncolor{latcolor}}c >{\columncolor{latcolor}}c}
		\toprule
		\multirow{3}{*}{\textbf{\makecell{LLM\\Backbone}}} & \multicolumn{6}{c|}{\textbf{Train-Ticket}} & \multicolumn{6}{c|}{\textbf{Online-Boutique}} & \multicolumn{6}{c}{\textbf{Simple-Micro}} \\
		\cmidrule(lr){2-7} \cmidrule(lr){8-13} \cmidrule(lr){14-19}
		& \multicolumn{3}{c}{\cellcolor{acccolor}\textit{Accuracy} (\%)} & \multicolumn{3}{c|}{\cellcolor{latcolor}\textit{Latency (s)}} & \multicolumn{3}{c}{\cellcolor{acccolor}\textit{Accuracy} (\%)} & \multicolumn{3}{c|}{\cellcolor{latcolor}\textit{Latency (s)}} & \multicolumn{3}{c}{\cellcolor{acccolor}\textit{Accuracy} (\%)} & \multicolumn{3}{c}{\cellcolor{latcolor}\textit{Latency (s)}} \\
		\cmidrule(lr){2-4} \cmidrule(lr){5-7} \cmidrule(lr){8-10} \cmidrule(lr){11-13} \cmidrule(lr){14-16} \cmidrule(lr){17-19} 
		& \textbf{\textit{Easy}} & \textbf{\textit{Med}} & \textbf{\textit{Hard}} & \textbf{\textit{Easy}} & \textbf{\textit{Med}} & \textbf{\textit{Hard}} & \textbf{\textit{Easy}} & \textbf{\textit{Med}} & \textbf{\textit{Hard}} & \textbf{\textit{Easy}} & \textbf{\textit{Med}} & \textbf{\textit{Hard}} & \textbf{\textit{Easy}} & \textbf{\textit{Med}} & \textbf{\textit{Hard}} & \textbf{\textit{Easy}} & \textbf{\textit{Med}} & \textbf{\textit{Hard}} \\
		\midrule
		
		\multicolumn{19}{c}{\cellcolor{blue!5} \textbf{\textit{Closed-Sourced LLMs}}} \\
		\midrule
		Qwen3-Plus & \underline{47.83} & 30.61 & 31.17 & 79.83 & 77.44 & 81.22 & \underline{43.48} & \underline{43.75} & 31.58 & 53.35 & 83.77 & 59.35 & \underline{47.83} & 36.17 & 38.03 & 71.26 & 75.79 & 78.18 \\
		Qwen3-Max & 47.83 & 28.57 & 30.77 & 69.87 & 86.38 & \textbf{12.36} & 39.13 & 37.50 & 25.32 & \underline{34.8}2 & 46.49 & 48.20 & 30.43 & 22.92 & 17.91 & 37.02 & 39.73 & 52.43 \\
		Qwen3-Flash & 21.74 & 16.33 & 13.16 & 43.24 & 84.96 & 290.7 & 34.78 & 30.61 & 21.33 & 42.33 & 77.26 & 98.64 & 22.72 & 14.58 & 8.86 & 50.52 & 61.12 & 65.33 \\
		
		\midrule
		\multicolumn{19}{c}{\cellcolor{cyan!5} \textbf{\textit{Open-Sourced LLMs}}} \\
		\midrule
		QwQ-32B & 17.39 & 10.20 & 7.89 & 157.4 & 194.1 & 183.2 & 26.09 & 22.45 & 15.58 & 109.9 & 137.3 & 155.5 & 17.39 & 8.33 & 6.76 & 141.5 & 188.7 & 195.7 \\
		Qwen3-Next & 13.04 & 6.12 & 5.06 & \textbf{23.33} & \textbf{20.41} & \underline{29.76} & 17.39 & 17.02 & 17.72 & \textbf{22.57} & \textbf{22.73} & \textbf{26.64} & 21.74 & 28.57 & 19.35 & \textbf{24.83} & \textbf{34.58} & \textbf{32.83} \\
		Qwen3-235B & 39.13 & \underline{34.69} & \underline{33.78} & 83.34 & 92.20 & 73.54 & 39.13 & 34.69 & 33.33 & 74.57 & 55.52 & 74.27 & 34.78 & 36.73 & 32.39 & 82.49 & 66.65 & 73.20 \\
		DeepSeek-V3.2 & 8.70 & 16.33 & 11.54 & 148.3 & 155.1 & 121.5 & 31.82 & 21.28 & 22.78 & 63.1 & 63.1 & 60.1 & 21.74 & 29.17 & 20.00 & 129.6 & 106.4 & 98.6 \\
		Kimi-K2 & 21.74 & 20.00 & 29.49 & 90.84 & 81.56 & 101.1 & 22.73 & 26.53 & 30.38 & 75.87 & 79.07 & 83.85 & 47.83 & \underline{44.89} & \underline{43.75} & 90.82 & 79.28 & 76.84 \\
		GLM-4.5 & 21.74 & 20.41 & 27.63 & 189.2 & 112.1 & 108.1 & 43.48 & 43.75 & \underline{39.47} & 135.1 & 126.1 & 130.8 & 43.48 & 36.73 & 30.38 & 127.8 & 126.3 & 132.8 \\
		
		\midrule
		\textit{E2E-REME} & \textbf{82.61} & \textbf{77.78} & \textbf{70.83} & \underline{39.65} & \underline{42.33} & 41.52 & \textbf{95.65} & \textbf{89.80} & \textbf{78.75} & 39.44 & \underline{38.31} & \underline{40.56} & \textbf{91.30} & \textbf{87.50} & \textbf{72.15} & \underline{49.65} & \underline{53.42} & \underline{55.33} \\
		\bottomrule
	\end{tabular}
	\vspace{0.5em}
	\caption{Remediation Accuracy (left columns) and Latency (right columns) across closed-source and open-source LLM backbones}
	\label{tab: accuracy_latency}
\end{table*}

\subsubsection{Reality-Anchored RFT}

During real-world deployment, each encountered failure instance $f_{\text{real}}$ is first handled by E2E-REME, producing a model-generated remediation playbook $a_{\text{real}}^{-}$. If the model-generated action does not successfully remediate the system, the failure is escalated to human operators. Site Reliability Engineers (SREs) then provide a corrective playbook $a_{\text{real}}^{+}$, which reflects an expert-preferred remediation strategy under the same conditions. This naturally yields a pairwise preference signal: $a_{\text{real}}^{+} \succ a_{\text{real}}^{-}$, indicating that the SRE action should be preferred over the model’s attempt.

Rather than relying on manually engineered scalar rewards or heuristic scoring functions, Reality-Anchored RFT employs \textit{Direct Preference Optimization} (DPO) to directly align E2E-REME with the remediation preferences demonstrated by SREs. Formally, let $\pi_\theta$ denote the policy of E2E-REME. The DPO objective encourages the model to increase the likelihood of generating the expert-preferred action $a_{\text{real}}^{+}$ while decreasing the likelihood of the model-generated action $a_{\text{real}}^{-}$.

\begin{equation}
	\left\{
	\begin{aligned}
		\mathcal{L}_{\text{DPO}}
		&= -\log \sigma\!\left(
		\beta\!\left[
		\Delta_\theta(a_{\text{real}}^{+}) - \Delta_\theta(a_{\text{real}}^{-})
		\right]\right)
		\\[4pt]
		\Delta_\theta(a) &\triangleq 
		\log\pi_\theta(a\mid f_{\text{real}})
		- \log\pi_{\text{ref}}(a\mid f_{\text{real}})
	\end{aligned}
	\right.
	\label{eq:reality-dpo}
\end{equation}

Formally, this preference-driven learning process is captured in Equation \ref{eq:reality-dpo}. Here, $\pi_{\text{ref}}$ denotes a frozen reference policy (typically the model obtained after Simulation-Based RFT), $\beta$ controls the sharpness of preference separation, and $\sigma(\cdot)$ is the logistic sigmoid function.

This stage enables E2E-REME to continuously internalize human expertise, ensuring that the model not only improves after deployment but also conforms to real-world safety conventions, operational best practices, and implicit SRE decision criteria.

\section{Evaluation}

In this section, we first introduce the implementation of E2E-REME, followed by the experimental setup. We then evaluate E2E-REME in terms of the following four research questions:

\begin{itemize}[leftmargin=*]
	\item \textbf{RQ1:} How accurately does E2E-REME perform microservice remediation compared to baseline LLMs?
	\item \textbf{RQ2:} What is the inference efficiency of E2E-REME—measured by runtime and token consumption?
	\item \textbf{RQ3:} How does each component in E2E-REME contribute to the final remediation accuracy?
	\item \textbf{RQ4:} How well does E2E-REME perform under realistic industrial workloads and microservice environments?
\end{itemize}

\subsection{Implementation \& Setting}

\subsubsection{Implementation}

We implement our algorithm using AgentEvolver~\cite{zhai2025agentevolver}, a self-evolving agent reinforcement-learning framework proposed by Alibaba Tongyi Lab. Unless otherwise stated, we set the training reward parameters to $\alpha = 1$, $\beta = 0.1$, $\gamma = 0.1$, $\delta = 0.5$, and $\lambda = 2$. The maximum retry number of ThinkRemed is set to $T_{\text{max}} = 1$. We adopt Qwen3-8B as the backbone LLM for all fine-tuning stages.

\subsubsection{Experimental Setup}

E2E-REME is trained on a CentOS 8 server equipped with 24 Intel(R) Xeon(R) CPUs (2.90GHz), 400GB RAM, and four NVIDIA A800 GPUs, each with 80GB of memory. The MicroRemed benchmark microservices are deployed across three machines, each equipped with 16 Intel(R) Xeon(R) CPUs (2.50GHz) and 64GB RAM.

To comprehensively evaluate the end-to-end microservice remediation capability of current LLMs, we examine a total of nine representative models, encompassing both closed-source and open-source variants. For fairness and consistency, all LLMs are executed within the ThinkRemed framework.

\textit{Closed-Source LLMs:} Qwen3-Plus, Qwen3-Max, and Qwen3-Flash~\cite{yang2025qwen3}.

\textit{Open-Source LLMs:} QwQ-32B, Qwen3-Next-80B-A3V, Qwen3-235B-A22B, DeepSeek-V3.2-Exp~\cite{liu2024deepseek}, Kimi-K2~\cite{team2025kimi}, and GLM-4.5~\cite{zeng2025glm}.

\subsection{RQ1: Remediation Accuracy}

We first compare the remediation accuracy of E2E-REME against other LLMs on the E2E-MR task. As illustrated in Table~\ref{tab: accuracy_latency}, the light-blue columns on the left summarize the accuracy under the easy, medium, and hard level settings for each microservice environment.

Overall, except for E2E-REME, Qwen3-Plus achieves the strongest performance among all evaluated models, followed by Qwen3-235B. At the microservice level, Train-Ticket emerges as the most challenging environment, followed by Simple-Micro. Notably, even under the easiest difficulty level, all standalone LLMs fail to exceed 50\% accuracy, underscoring the difficulty and rigor of the MicroRemed benchmark.

In contrast, E2E-REME consistently outperforms the best baseline, Qwen3-Plus, by 56.48\%, 48.50\%, and 42.97\% across the three microservice environments, demonstrating its clear superiority in end-to-end microservice remediation.

\subsection{RQ2: Inference Efficiency}

We next compare the remediation latency of E2E-REME with other LLMs. As shown in the light-gray columns of Table~\ref{tab: accuracy_latency}, the latency reflects the end-to-end time of a full remediation cycle, including model reasoning, system probing, action execution, and recovery verification.

\begin{figure}[htbp]
	\centering
	\includegraphics[width=1\linewidth]{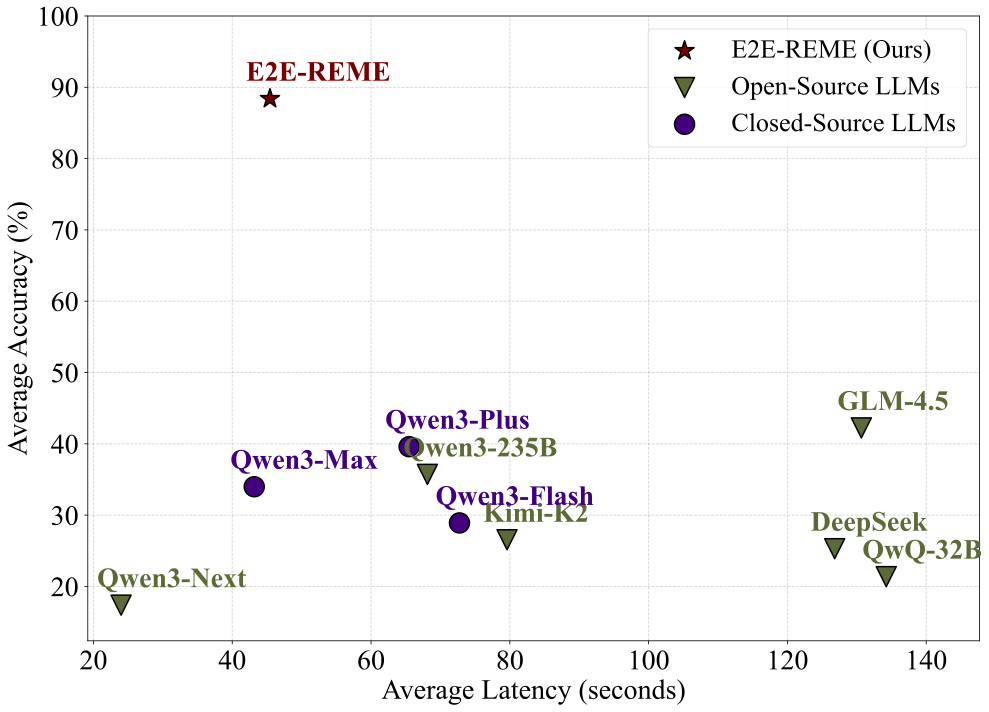}
	\caption{Latency–accuracy trade-off of various large language models on the Online-Boutique microservice}
	\label{fig: latency-accuracy-capacity}
\end{figure}

Overall, Qwen3-Next exhibits consistently low latency across all environments, indicating a lightweight reasoning pipeline and efficient prompt handling. However, when considered together with its accuracy, this speed advantage comes at the expense of insufficient reasoning depth and unstable remediation performance, rendering it largely impractical for real-world use. In contrast, E2E-REME achieves the second-lowest latency among all evaluated models—incurring only modest overhead compared to Qwen3-Next (higher by 40.48\%, 39.19\%, and 41.76\% across the three microservices, respectively), while remaining substantially faster than Qwen3-Flash (lower by 70.52\%, 45.79\%, and 10.49\%). More importantly, E2E-REME delivers consistently high accuracy, striking a favorable balance between efficiency and effectiveness and making it suitable for practical end-to-end microservice remediation.

\begin{table}[h]
	\setlength{\tabcolsep}{2pt}
	\centering
	\begin{tabular}{c|ccc|ccc}
		\toprule
		\multirow{2}{*}{\textbf{Backbone}} &\multicolumn{3}{c|}{Train-Ticket} & \multicolumn{3}{c}{Online-Boutique} \\
		\\[-2.5ex]
		\cline{2-7}
		\\[-2ex]
		~ & \textit{Easy} & \textit{Medium} & \textit{Hard} & \textit{Easy} & \textit{Medium} & \textit{Hard} \\
		\midrule
		\multicolumn{7}{c}{\cellcolor{brightlavender!15} \textbf{\textit{Closed-Sourced LLMs}}} \\
		\midrule
		Qwen3-Plus & 3363 & 4988 & 4359 & 18822 & 127299 & 108053\\
		Qwen3-Max & 3583 & \underline{2145} & \underline{3256} & 4821 & 5366 & 5758 \\
		Qwen3-Flash & \underline{2948} & 3651 & 3891 & 3490 & \underline{3003} & 3378 \\
		\midrule			
		\multicolumn{7}{c}{\cellcolor{capri!15} \textbf{\textit{Open-Sourced LLMs}}} \\
		\midrule
		QwQ-32B & 4147 & 5918 & 5936 & \underline{2003} & 3784 & \underline{2951} \\			
		Qwen3-Next & 14490 & 13267 & 13091 & 12190 & 9387 & 24385 \\		
		Qwen3-235B & 4858 & 8369 & 6669 & 10624 & 15749 & 30381\\								
		DeepSeek-V3.2 & 5195 & 7295 & 7785 & 5855 & 6004 & 6454  \\	
		Kimi-K2 & 6452 & 4964 & 6728 & 4810 & 6314 & 7793 \\
		GLM-4.5 & 11264 & 9492 & 10652 & 11270 & 11991 & 10692 \\
		\midrule
		E2E-REME(\textit{ours}) & \textbf{1715} & \textbf{1844} & \textbf{1735} & \textbf{1552} & \textbf{2223} & \textbf{1648} \\
		\bottomrule
	\end{tabular}
	\vspace{0.5em}
	\caption{Average Token Consumption per remediation results}
	\label{tab: more-token-consumption}
	\vspace{-2em}
\end{table}

To provide a clearer comparison, we further plot the latency–accuracy trade-off in Figure~\ref{fig: latency-accuracy-capacity}, where both latency and accuracy are averaged over the three difficulty levels (Easy, Medium, and Hard) on the Online-Boutique microservice. Each point represents a model, where the x-axis denotes average inference latency (lower is better) and the y-axis indicates accuracy. The plot highlights the superiority of E2E-REME, which lies closest to the upper-left region. Relative to the second-best model, Qwen3-Max, E2E-REME achieves 57.71\% higher accuracy while reducing latency by 8.65\%, respectively.

We further compare the average token consumption of each model, as shown in Table~\ref{tab: more-token-consumption}. The reported values include both input and output tokens, thereby reflecting the total reasoning and generation workload for each remediation process. The results align with the latency findings, though a few exceptions provide additional insight. For example, Qwen3-Plus and Qwen3-Next consume significantly more tokens without proportional increases in latency—mainly due to unnecessary probing steps that generate overly long command outputs. Aside from these cases, the results further corroborate the efficiency of E2E-REME: compared with the second most efficient model, Qwen3-Flash, its token consumption is lower by 49.53\% and 45.06\%, respectively.

\subsection{RQ3: Ablation Study}

To assess the contribution of each training stage in E2E-REME, we conduct an ablation study on the Online-Boutique microservice. The results are summarized in Table \ref{tab:ablation-study-training}. Starting from the raw Qwen3-8B backbone, the model achieves only 30.43 \%, 20.83\%, and 15.56\% accuracy across the three difficulty levels, confirming that the base model lacks the specialized knowledge required for effective end-to-end remediation.

\begin{table}[h]
	\setlength{\tabcolsep}{3.0pt}
	\centering
	\begin{tabular}{l | >{\columncolor{acccolor}}c >{\columncolor{acccolor}}c >{\columncolor{acccolor}}c| >{\columncolor{latcolor}}c >{\columncolor{latcolor}}c >{\columncolor{latcolor}}c}
		\toprule
		\multicolumn{1}{c|}{\multirow{2}{*}{\textit{Method}}} & \multicolumn{3}{c|}{\cellcolor{acccolor}\textit{Accuracy} (\%)} & \multicolumn{3}{c}{\cellcolor{latcolor}\textit{Latency (s)}} \\
		\\[-2.5ex]
		\cline{2-7}
		\\[-2ex]
		~ & \textit{Easy} & \textit{Medium} & \textit{Hard} & \textit{Easy} & \textit{Medium} & \textit{Hard} \\
		\midrule
		Qwen3-8B & 30.43 & 20.83 & 15.56 & 359.35 & 319.24 & 343.13 \\
		\textbf{+}SFT & 34.78 & 24.49 & 22.50 & 263.72 & 250.08 & 285.43 \\
		\quad \textbf{+}Sim-RFT & \underline{86.96} & \underline{85.71} & \underline{69.62} & \underline{36.37} & \underline{50.47} & \underline{51.84} \\
		\quad  \quad \textbf{+}Real-RFT & \textbf{95.65} & \textbf{89.80} & \textbf{78.75} & \textbf{39.44} & \textbf{38.31} & \textbf{40.56} \\
		\bottomrule
	\end{tabular}
	\vspace{0.5em}
	\caption{Ablation study on the Online-Boutique microservice}
	\label{tab:ablation-study-training}
	\vspace{-2em}
\end{table}

Introducing SFT yields a substantial improvement, boosting accuracy by 3.66\%–6.94\% across different levels. This demonstrates that supervised instruction tuning on high-quality remediation trajectories provides essential task grounding and significantly enhances the model’s action planning ability, while also reducing inference latency by 21.78\% due to fewer unnecessary probing steps.

Adding Sim-RFT yields the largest performance gain, delivering an additional 53.61\% improvement. The gains indicate that synthetic experience simulation effectively enriches the model's exposure to diverse failure-recovery patterns, enabling more robust reasoning in unseen scenarios. Finally, Real-RFT provides an additional but smaller improvement of +7.24\% on accuracy. The modest gain is primarily because the failure types and microservices used in Real-RFT training differ from those in the Online-Boutique evaluation environment. Nevertheless, even with this mismatch, Real-RFT still contributes to more consistent decision-making and a small reduction in execution time (-14.69\%).

We further validate the effectiveness of each component in ThinkRemed. To isolate the impact of our training pipeline, we perform this analysis using Qwen3-Plus—the strongest LLM backbone aside from E2E-REME—instead of our trained model.

\begin{table}[h]
	\setlength{\tabcolsep}{3.1pt}
	\centering
	\begin{tabular}{c|ccc|ccc}
		\toprule
		\multicolumn{1}{c|}{\multirow{2}{*}{\textit{Method}}} & \multicolumn{3}{c|}{\textbf{Train-Ticket}} & \multicolumn{3}{c}{\textbf{Online-Boutique}} \\
		\\[-2.5ex]
		\cline{2-7}
		\\[-2ex]
		~ & \textit{Easy} & \textit{Medium} & \textit{Hard} & \textit{Easy} & \textit{Medium} & \textit{Hard} \\
		\midrule
		\textbf{ThinkRemed} & \textbf{47.83} & 30.61 & \textbf{31.17} & \textbf{43.48} & \textbf{43.75} & \textbf{31.58} \\
		\midrule
		w/o Probe & \underline{43.48} & \textbf{34.69} & \underline{30.38} & \underline{39.13} & \underline{40.43} & \underline{30.38} \\
		w/o Reflection & \underline{43.48} & 28.57 & 26.92 & 34.78 & 36.17 & 25.32 \\
		w/o P. \& R. & 39.13 & \underline{33.33} & 20.51 & 30.43 & 35.42 & 20.51  \\
		\bottomrule
	\end{tabular}
	\vspace{0.5em}
	\caption{ThinkRemed's ablation study}
	\label{tab:ablation-study-thinkremed}
	\vspace{-2em}
\end{table}

As shown in Table~\ref{tab:ablation-study-thinkremed}, we evaluate three variants: removing the probe agent, removing reflection, and removing both. Overall, both components contribute positively. For example, in the Train-Ticket microservice (easy level), removing either probe or reflection reduces accuracy by 13.05\%. Across all settings, reflection plays a more critical role than probing: removing reflection results in an average 5.53\% accuracy drop, whereas removing the probe agent decreases accuracy by only 1.66\%.

\subsection{RQ4: Industrial Evaluation}

To assess the practical effectiveness of E2E-REME under realistic industrial workloads and microservice environments, we conduct an evaluation on three production microservices paired with their real operational traffic. Unlike the benchmark experiments—which directly measure accuracy and latency—this study focuses on remediation time reduction, a metric that more faithfully captures the efficiency gains experienced by Site Reliability Engineers (SREs). For each incident, SREs manually recorded the time required to complete remediation both with and without E2E-REME, and the relative reduction was used as the final metric.

\begin{table}[h]
	\setlength{\tabcolsep}{3.0pt}
	\centering
	\begin{tabular}{c|ccc}
		\toprule
		\multirow{2}{*}{\textbf{Method}} & \multicolumn{3}{c}{\textbf{Remediation Time Reduction (\%)}} \\ 
		\cmidrule(lr){2-4} 
		~ & \textit{Microservice A} & \textit{Microservice B} & \textit{Microservice C} \\
		\midrule
		MAPE-Ansible & 67.21 & 55.83 & 48.95 \\
		WCA-Ansible & 73.76 & 60.90 & 56.74 \\
		E2E-REME(\textit{ours}) & \textbf{81.33} & \textbf{79.58} & \textbf{76.03} \\ 
		\bottomrule
	\end{tabular}
	\vspace{0.5em}
	\caption{Evaluation on 3 realistic industrial workloads and microservices: Percentage reduction in remediation time}
	\label{tab:industrial-evaluation}
	\vspace{-2em}
\end{table}

We compare E2E-REME against two representative industrial automation systems, MAPE-Ansible~\cite{sarda2024leveraging} and WCA-Ansible~\cite{sahoo2024ansible}. As shown in Table~\ref{tab:industrial-evaluation}, E2E-REME consistently achieves the largest reduction in remediation time across all three microservices. While MAPE-Ansible (powered by GPT-5) achieves reductions between 48.95\% and 67.21\%, and WCA-Ansible provides moderate improvements of up to 73.76\%, E2E-REME delivers substantially stronger reductions—improving over the second-best WCA-Ansible by 7.57\%, 18.68\%, and 19.29\%, respectively.

These findings indicate that even without being explicitly fine-tuned on the target industrial system, E2E-REME can generate robust and reliable remediation strategies that meaningfully accelerate human operational workflows. This demonstrates the method’s strong potential for practical adoption in production environments. That said, fully realizing the vision of end-to-end microservice remediation—ideally reducing SRE intervention by 99\%—will require future work on broader environment robustness, safety mechanisms, and tighter integration with production automation pipelines.

\section{Threats to Validity}

We discuss the limitations of our work from two perspectives: the benchmark and the methodology.

\noindent \textbf{Benchmark.}
Although the MicroRemed benchmark provides sufficient challenges for evaluating end-to-end microservice remediation, the currently supported failure types remain limited—covering only seven of the most common categories. In real-world systems, failure modes are far more diverse and continuously evolving~\cite{zhang2024multivariate, zhang2024towards, wang2025survey}. Nevertheless, the design of MicroRemed inherently supports extensibility; new failure types can be integrated seamlessly. The main challenge lies in the need to implement corresponding fault injection and detection mechanisms when introducing additional failure types.

\noindent \textbf{Methodology.}
While E2E-REME demonstrates strong performance on the end-to-end microservice auto-remediation task, its long-term stability and generalization still warrant further investigation. Although we include evaluations in realistic industrial environments, the inherent unpredictability of LLMs means that there is always a non-negligible risk of generating incorrect actions that could destabilize or even break the cluster. Therefore, fully deploying such systems in production requires additional safeguard mechanisms—such as action verification, safety filters, or fail-safe rejection modules—to ensure robust and reliable operation.

\section{Related Work}

\subsection{Software Remediation}

Software remediation, as the next step beyond failure diagnosis, has long been studied as a generation problem. Existing work can be broadly grouped into two categories: mitigation solution generation and remediation script generation.

\noindent \textbf{Mitigation Solution Generation.}
These approaches focus on generating actionable mitigation strategies for detected anomalies, often leveraging large corpora of historical incident reports. Toufique et al.~\cite{ahmed2023recommending} conduct the first large-scale study evaluating LLMs for root-cause analysis and mitigation in production incidents. Drishti et al.~\cite{goel2024x} integrate signals from the entire software development lifecycle and apply retrieval-augmented in-context learning to enhance mitigation quality. Pouya et al.~\cite{hamadanian2023holistic} model the natural workflow of on-call engineers using three LLM agents—hypothesis formation, hypothesis testing, and mitigation planning.

\noindent \textbf{Remediation Script Generation.}
These methods aim to produce executable scripts or code snippets that directly automate repair actions. Xpert~\cite{jiang2024xpert} generates tailored KQL queries for incident investigation. ShellGPT~\cite{shi2023shellgpt} fine-tunes GPT models for shell command recommendation. Wisdom-Ansible~\cite{pujar2023automated} and MAPE-Ansible~\cite{sarda2024leveraging} generate Ansible playbooks using fine-tuned or GPT-4-based MAPE-K architectures. WCA-Ansible~\cite{sahoo2024ansible} further pretrains a domain-specific model on natural language, source code, and Ansible data to enhance playbook generation. Our work falls into this category, but differs in aiming for end-to-end microservice remediation without relying on human-written playbooks or static domain knowledge.

\subsection{LLM-based Failure Management}

Large language models have recently been applied to enhance anomaly detection, failure diagnosis, and automated remediation in complex systems~\cite{zhang2025survey, zhang2025scalalog}. Existing efforts can be broadly grouped into foundation models, fine-tuning-based approaches, and prompt-driven methods.

\noindent \textbf{Foundation models for system data.}
A number of works aim to build LLM-style foundation models for time-series or log data. Representative examples include Lag-Llama~\cite{rasul2023lag}, Timer~\cite{liu2024timer}, and TimesFM~\cite{das2024decoder}, which unify forecasting, imputation, and detection under transformer architectures. PreLog~\cite{le2024prelog} and KAD-Disformer~\cite{yu2024pre} extend this line to log parsing and multivariate anomaly detection. These models provide domain-specialized priors but are typically limited to a single modality.

\noindent \textbf{Fine-tuning general LLMs.}
Another line of work fine-tunes general-purpose LLMs for IT operations. Examples include UniTime~\cite{liu2024unitime}, AnomalyLLM~\cite{liu2024anomalyllm}, and LogLM~\cite{liu2024loglm}, which adapt GPT/LLaMA-style architectures for time-series forecasting, anomaly detection, and log analytics. RAG4ITOps~\cite{zhang2024rag4itops} and OWL~\cite{guo2023owl} further incorporate retrieval or adapter tuning for interactive diagnosis. These methods improve task specialization, but require extensive labeled data or careful adaptation. Our E2E-REME also falls into this category, but differs by integrating multi-stage reinforcement fine-tuning for end-to-end remediation.

\noindent \textbf{Prompt-based methods.}
Prompt-based solutions avoid heavy fine-tuning by leveraging instruction design, chain-of-thought reasoning, or retrieval. RCACopilot~\cite{chen2024automatic}, Xpert~\cite{jiang2024xpert}, and LasRCA~\cite{han2024potential} design multi-step prompts for diagnosis and anomaly detection. LogGPT~\cite{liu2024interpretable}, LSTPrompt~\cite{liu2024lstprompt}, and LM-PACE~\cite{zhang2024lm} improve interpretability through CoT or task decomposition. Retrieval-augmented approaches such as RAGLog~\cite{pan2023raglog}, LogRAG~\cite{zhang2024lograg} and XRAGLog~\cite{zhang2025xraglog} enhance log reasoning through case retrieval. While flexible, prompt-based methods often suffer from unstable generation and limited end-to-end automation.

\section{Conclusion}

In this paper, we introduce the task of end-to-end microservice remediation. To enable systematic evaluation, we construct MicroRemed, a challenging benchmark that automates microservice deployment, failure injection, playbook execution, and post-repair verification. To tackle this task, we propose E2E-REME, an end-to-end auto-remediation model for microservices based on experience-simulation reinforcement fine-tuning. Experimental results show that MicroRemed presents substantial challenges for existing LLMs, while E2E-REME achieves superior accuracy and efficiency on both MicroRemed and realistic industrial microservice environments.

\begin{acks}
	This work is supported by Key RD Project of Guangdong Province, China (No.2020B010164003).
\end{acks}

\bibliographystyle{ACM-Reference-Format}
\balance
\bibliography{sample-base}

\clearpage

\end{sloppypar}
\end{document}